\documentclass[runningheads,citeauthoryear]{apinv}
\usepackage{epsfig,cite,graphics}

\begin{document}

\title{Multiband optical surface brightness profile decompositions of the
Seyfert galaxies \newline Mrk~79 and NGC~5548\thanks{Based on observations obtained at
the Rozhen National Astronomical Observatory of Bulgaria operated by the
Institute of Astronomy, Bulgarian Academy of Sciences.}}
\titlerunning{Surface brightness profiles decomposition of Mrk~79 and NGC~5548}
\author{Boyko Mihov, Lyuba Slavcheva-Mihova}
\authorrunning{B. Mihov, L. Slavcheva-Mihova}
\tocauthor{Boyko M. Mihov}
\institute{Institute of Astronomy, Bulgarian Academy of Sciences, \newline
72 Tsarigradsko Chausse Blvd., Sofia 1784, Bulgaria \newline
\email{bmihov@astro.bas.bg}}

\papertype{conference poster}
\maketitle
\begin{abstract}
We present preliminary results of the Johnson-Cousins $(U)BVRI$ surface brightness profile
decompositions of the Seyfert galaxies Mrk~79 and NGC~5548. The profiles were modelled
as a sum of a Gaussian law for the nucleus, a S\'ersic law for the bulge and an exponent law
for the disk. A flat bar was added to the model profile of Mrk~79. The parameters and
the total magnitudes of the structural components were derived.
\end{abstract}

\section*{Introduction}
We present the preliminary decomposition results for the Johnson-Cousins
$(U)BVRI$ surface brightness profiles of Seyfert galaxies. Observations and data reduction were
described in Slavcheva-Mihova et~al.\cite{sm05}, the decomposition procedure
was presented in Slavcheva-Mihova et~al.\cite{sm06}. The data were obtained at the
Rozhen National Astronomical Observatory of Bulgaria using the 2.0-m telescope and the Photometrics
AT200 model CCD camera. Data reduction was done via ESO-MIDAS package and profiles were extracted by means
of the modified version of FIT/ELL3 ellipse fitting command. Decompositions were done using unweighted
non-linear least-squares method for the simultaneous fitting of the composite model profile to the observed
one. The model profiles library includes a Gaussian law for nuclei and rings, a S\'ersic law for bulges,
pure and truncated exponent laws for disks and a flat bar profile for bars (see Prieto et~al. \cite{p97};
Prieto et~al. \cite{p01}; Aguerri et~al. \cite{a05}). Spiral arms appear as bumps onto the disk-dominated
part of surface brightness profiles. The data points corresponding to them were excluded from the fit through
assigning zero weights. Different combinations of model profiles were used for the individual galaxies in order
to get satisfactory fits to the observational profiles.

\section{Results}
\subsection{Mrk~79}
\begin{figure}[t]
\begin{center}
\includegraphics[scale=0.66]{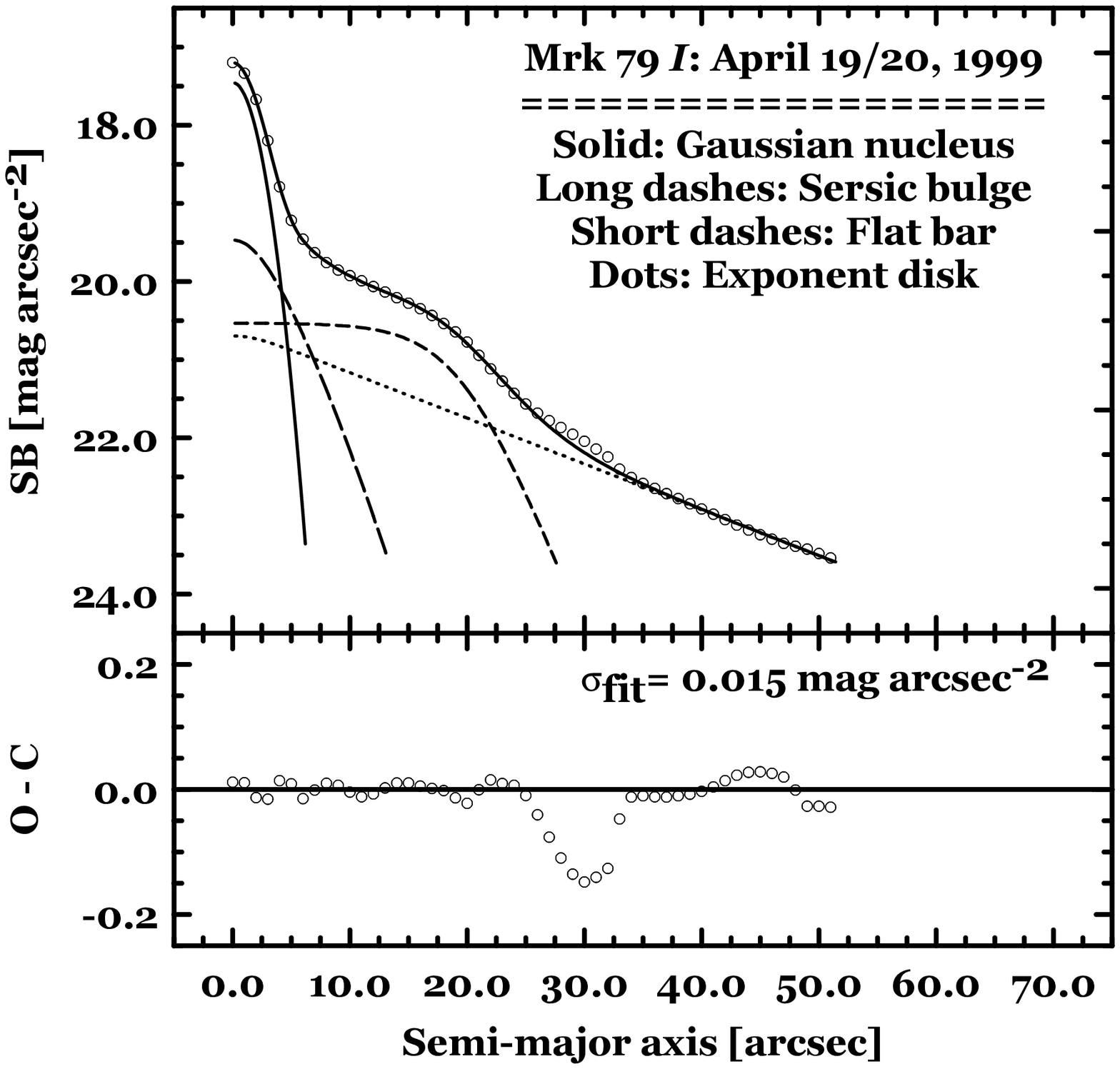}
\caption[]{$I$ band surface brightness profile decomposition for Mrk~79. The open circle line is the
observed profile and the solid line is the model one. Observed minus calculated profile, O~--~C, is also shown.}
\label{decres1}
\end{center}
\end{figure}

The surface brightness profiles of Mrk~79 were decomposed into a Gaussian nucleus,
a S\'ersic bulge, an exponent disk and a flat bar (see Fig.~\ref{decres1}). The fitted model
parameters and their errors are listed in Table~\ref{fitpar} and in Table~\ref{barpar};
in all tables the errors of the parameters are listed next to their values. The total
magnitudes of the structural components and their errors are listed in Table~\ref{totmag}
(total bar magnitude is not listed). Two regions of the profiles were excluded from the
fit~-- these regions reflect the galaxy two-arm spiral structure. The spiral arms start
from the bar end and for a half revolution pass close to the bar producing the first bump
of the surface brightness profiles around $a$~= 30 arcsec next to the bar end and then
open through the disk producing the second bump around $a$~= 40~--~45 arcsec.

\subsection{NGC~5548}
\begin{figure}[t]
\begin{center}
\includegraphics[scale=0.66]{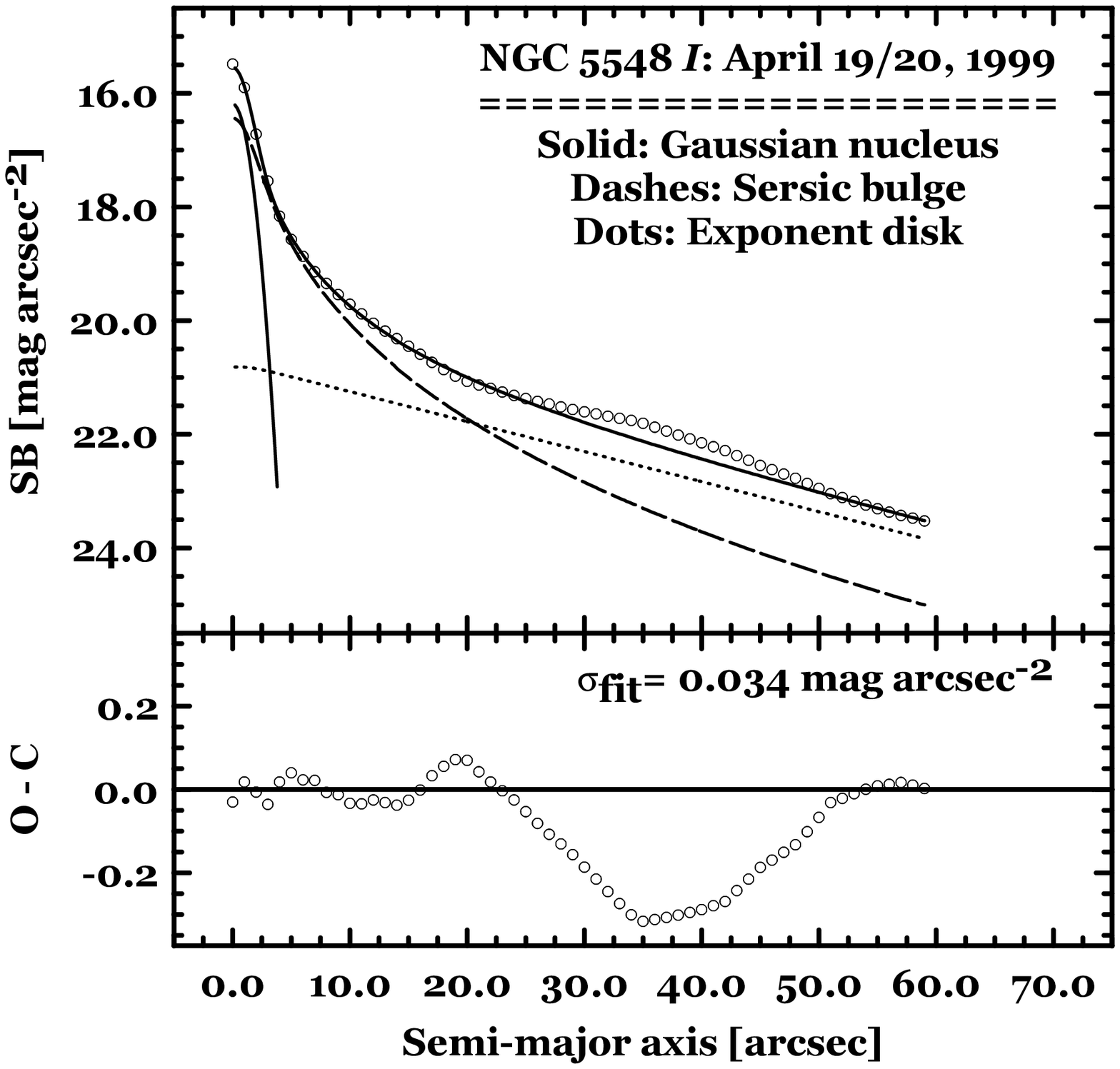}
\caption[]{The same as in Fig.~\ref{decres1}, but for NGC~5548.}
\label{decres2}
\end{center}
\end{figure}

The surface brightness profiles of NGC~5548 were decomposed into a Gaussian nucleus,
a S\'ersic bulge and an exponent disk (see Fig.~\ref{decres2}). The region of the surface
brightness profiles around $a$~= 35 arcsec is influenced by the galaxy spiral arms and
was excluded from the fit. The fitted model parameters and their errors are listed in
Table~\ref{fitpar}. The total magnitudes of the structural components and their errors
are listed in Table~\ref{totmag}.

\begin{table}[t]
\begin{center}
\caption{Fitted parameters of the bulge (effective surface brightness, $\mu_{\rm eff}$, effective radius, $r_{\rm eff}$,
and S\'ersic power-law index, $n$) and of the disk (central surface brightness, $\mu_{0}$, and scale length, $r_{\rm scl}$)
components of Mrk~79 and NGC~5548.}
\begin{tabular}{@{}l@{\hspace{0.35cm}}c@{\hspace{0.45cm}}c@{\hspace{0.45cm}}c@{\hspace{0.45cm}}c@{\hspace{0.45cm}}c@{}}
Galaxy/Band/Epoch & $\mu_{\rm eff}$ & $r_{\rm eff}$ & $n$ & $\mu_0$ & $r_{\rm scl}$ \\
           & [$\rm mag\,\rm arcsec^{-2}$] & [arcsec] & $\cdots$ & [$\rm mag\,\rm arcsec^{-2}$] & [arcsec] \\
\noalign{\smallskip} \hline
\noalign{\smallskip}
Mrk~79/$U$/Apr.'99 & $\cdots$     & $\cdots$     & $\cdots$    & 21.781 0.064 & 12.318 0.240 \\
Mrk~79/$B$/Feb.'99 & $\cdots$     & $\cdots$     & $\cdots$    & 21.604 0.025 & 13.194 0.077 \\
Mrk~79/$B$/Apr.'99 & 21.987 1.204 &  3.363 1.509 & 1.150 1.158 & 22.369 0.012 & 18.160 0.071 \\
Mrk~79/$V$/Feb.'99 & 18.530 0.578 &  1.326 0.226 & 1.994 0.193 & 21.143 0.013 & 14.582 0.051 \\
Mrk~79/$V$/Apr.'99 & 19.522 0.491 &  1.981 0.359 & 3.107 0.329 & 21.504 0.012 & 18.087 0.075 \\
Mrk~79/$R$/Feb.'99 & 19.528 1.564 &  2.463 1.445 & 1.659 1.553 & 20.911 0.021 & 16.380 0.093 \\
Mrk~79/$R$/Apr.'99 & 21.243 0.242 &  5.777 0.967 & 1.377 0.552 & 21.086 0.051 & 17.998 0.267 \\
Mrk~79/$I$/Feb.'99 & 22.420 0.684 & 14.666 5.144 & 7.637 2.333 & 20.262 0.027 & 15.990 0.334 \\
Mrk~79/$I$/Apr.'99 & 20.403 0.057 &  5.100 0.235 & 0.747 0.154 & 20.578 0.021 & 18.527 0.143 \\
\noalign{\smallskip} \hline
\noalign{\smallskip}
NGC~5548/$U$/Apr.'99 &      $\cdots$ &     $\cdots$ &    $\cdots$ &   $\cdots$   &   $\cdots$   \\
NGC~5548/$B$/Apr.'99 &  20.908 0.174 &  6.276 0.531 & 5.437 1.314 & 22.836 0.083 & 22.827 0.492 \\
NGC~5548/$V$/Apr.'99 &  21.008 0.551 &  9.161 2.672 & 4.564 1.273 & 22.138 0.254 & 19.580 1.026 \\
NGC~5548/$R$/Apr.'99 &  19.743 0.071 &  6.149 0.284 & 1.663 0.188 & 20.895 0.127 & 18.149 0.633 \\
NGC~5548/$I$/Apr.'99 &  19.753 0.307 &  8.573 1.452 & 3.701 0.558 & 20.721 0.169 & 20.537 0.260 \\
\end{tabular}
\label{fitpar}
\end{center}
\end{table} 

\begin{table}[t]
\begin{center}
\caption{Fitted parameters of the bar (central surface brightness, $\mu_{0}$, bar length, $r_{\rm bar}$, and scale length,
$r_{\rm scl}$) component of Mrk~79.}
\begin{tabular}{@{}l@{\hspace{0.35cm}}c@{\hspace{0.45cm}}c@{\hspace{0.45cm}}c@{}}
Galaxy/Band/Epoch & $\mu_0$ & $r_{\rm bar}$ & $r_{\rm scl}$ \\
 & [$\rm mag\,\rm arcsec^{-2}$] & [arcsec] & [arcsec] \\
\noalign{\smallskip} \hline
\noalign{\smallskip}
Mrk~79/$U$/Apr.'99 & 23.374 0.072 & 22.227 0.191 & 1.388 0.313 \\
Mrk~79/$B$/Feb.'99 & 23.239 0.041 & 21.884 0.114 & 1.376 0.138 \\
Mrk~79/$B$/Apr.'99 & 22.667 0.028 & 21.285 0.103 & 2.248 0.048 \\
Mrk~79/$V$/Feb.'99 & 22.095 0.016 & 20.287 0.069 & 2.034 0.045 \\
Mrk~79/$V$/Apr.'99 & 21.903 0.018 & 20.305 0.070 & 2.221 0.046 \\
Mrk~79/$R$/Feb.'99 & 21.263 0.044 & 19.738 0.167 & 2.495 0.086 \\
Mrk~79/$R$/Apr.'99 & 21.340 0.061 & 20.519 0.112 & 2.258 0.081 \\
Mrk~79/$I$/Feb.'99 & 21.016 0.026 & 19.768 0.064 & 2.114 0.046 \\
Mrk~79/$I$/Apr.'99 & 20.532 0.053 & 19.489 0.222 & 2.934 0.083 \\
\end{tabular}
\label{barpar}
\end{center}
\end{table} 

\begin{table}[t]
\begin{center}
\caption{Total magnitudes of the nucleus, bulge and disk components of Mrk~79 and NGC~5548.
The mean FWHM of the frame PSF, $\langle\cal {FW}_{\rm \scriptscriptstyle PSF}\rangle$,
and the standard deviation of the fit, $\sigma_{\rm fit}$, are also listed.}
\begin{tabular}{@{}l@{\hspace{0.35cm}}c@{\hspace{0.45cm}}c@{\hspace{0.45cm}}c@{\hspace{0.45cm}}c@{\hspace{0.45cm}}c@{}}
Galaxy/Band/Epoch & Nucleus & Bulge & Disk & $\langle\cal {FW}_{\rm \scriptscriptstyle PSF}\rangle$ & $\sigma_{\rm fit}$ \\
 & [mag] & [mag] & [mag] & [arcsec] & [$\rm mag\,\rm arcsec^{-2}$] \\
\noalign{\smallskip} \hline
\noalign{\smallskip}
Mrk~79/$U$/Apr.'99 & 14.139 0.013 &     $\cdots$ & 14.333 0.077 & 4.185 & 0.063 \\
Mrk~79/$B$/Feb.'99 & 14.747 0.012 &     $\cdots$ & 14.007 0.028 & 3.224 & 0.050 \\
Mrk~79/$B$/Apr.'99 & 15.368 0.194 & 16.594 1.549 & 14.078 0.015 & 3.975 & 0.017 \\
Mrk~79/$V$/Feb.'99 & 15.322 0.320 & 14.888 0.686 & 13.328 0.015 & 3.119 & 0.021 \\
Mrk~79/$V$/Apr.'99 & 15.853 0.319 & 14.781 0.629 & 13.222 0.015 & 3.687 & 0.016 \\
Mrk~79/$R$/Feb.'99 & 14.311 0.467 & 14.633 2.017 & 12.844 0.024 & 3.234 & 0.030 \\
Mrk~79/$R$/Apr.'99 & 14.340 0.045 & 14.589 0.437 & 12.814 0.060 & 3.205 & 0.017 \\
Mrk~79/$I$/Feb.'99 & 14.336 0.115 & 12.858 1.024 & 12.247 0.053 & 3.138 & 0.012 \\
Mrk~79/$I$/Apr.'99 & 14.085 0.026 & 14.304 0.115 & 12.244 0.027 & 3.494 & 0.015 \\
\noalign{\smallskip} \hline
\noalign{\smallskip}
NGC~5548/$U$/Apr.'99 &     $\cdots$ &     $\cdots$ &     $\cdots$ & 3.495 & $\cdots$ \\
NGC~5548/$B$/Apr.'99 & 14.170 0.106 & 13.370 0.253 & 14.048 0.095 & 2.906 & 0.042 \\
NGC~5548/$V$/Apr.'99 & 14.321 0.085 & 12.741 0.839 & 13.683 0.278 & 2.679 & 0.044 \\
NGC~5548/$R$/Apr.'99 & 13.397 0.032 & 12.860 0.123 & 12.605 0.148 & 2.470 & 0.036 \\
NGC~5548/$I$/Apr.'99 & 14.025 0.070 & 11.740 0.479 & 12.163 0.171 & 2.535 & 0.034 \\

\end{tabular}
\label{totmag}
\end{center}
\end{table} 

\end{document}